\documentclass[12pt]{iopart}

\usepackage{iopams}
\usepackage[T1]{fontenc}
\usepackage[utf8]{inputenc}
\usepackage[usenames,dvipsnames]{xcolor}
\usepackage{graphicx}
\usepackage{setstack}
\usepackage{multirow}
\usepackage{siunitx}
\usepackage{cite}

\begin{document}

\title[Preparing single SiV$^{-}$ center in nanodiamonds for external, optical coupling]{Preparing single SiV$^{-}$ center in nanodiamonds for external, optical coupling with access to all degrees of freedom}

\author{Stefan H\"au{\ss}ler \textsuperscript{\bfseries 1,2},
  Lukas Hartung \textsuperscript{\bfseries 1},
  Konstantin G. Fehler \textsuperscript{\bfseries 1,2},
  Lukas Antoniuk \textsuperscript{\bfseries 1},
  Liudmila F. Kulikova \textsuperscript{\bfseries 3},
  Valery A. Davydov \textsuperscript{\bfseries 3},
  Viatcheslav N. Agafonov \textsuperscript{\bfseries 4},
  Fedor Jelezko \textsuperscript{\bfseries 1,2} and
  Alexander Kubanek \textsuperscript{\bfseries 1,2}}

\address{\textsuperscript{1} Institute for Quantum Optics, Ulm University, Albert-Einstein-Allee 11, 89081 Ulm, Germany \\
  \textsuperscript{2} Center for Integrated Quantum Science and Technology (IQst),
Ulm University, Albert-Einstein-Allee 11, 89081 Ulm, Germany \\
  \textsuperscript{3} L.F. Vereshchagin Institute for High Pressure Physics,
Russian Academy of Sciences, Troitsk, Kaluzhskoe shosse 14, Moscow 142190, Russia \\
  \textsuperscript{4} GREMAN, UMR CNRS CEA 6157, Université F. Rabelais, Parc de Grandmont, 37200 Tours, France}
  
\ead{alexander.kubanek@uni-ulm.de}

\vspace{10pt}

\begin{indented}
\item[]October 2019
\end{indented}

\begin{abstract}

Optical coupling enables intermediate- and long-range interactions between distant quantum emitters. Such interaction may be the basic element in bottom-up approaches of coupled spin systems or for integrated quantum photonics and quantum plasmonics.
Here, we prepare nanodiamonds carrying single, negatively-charged silicon-vacancy centers for evanescent optical coupling with access to all degrees of freedom by means of atomic force nanomanipulation. The color centers feature excellent optical properties, comparable to silicon-vacancy centers in bulk diamond, resulting in a resolvable fine structure splitting, a linewidth close to the Fourier-Transform limit under resonant excitation and a good polarization contrast. We determine the orbital relaxation time $T_{1}$ of the orbitally split ground states and show that all optical properties are conserved during translational nanomanipulation. Furthermore, we demonstrate the rotation of the nanodiamonds. In contrast to the translational operation, the rotation leads to a change in polarization contrast. We utilize the change in polarization contrast before and after nanomanipulation to determine the rotation angle. Finally, we evaluate the likelihood for indistinguishable, single photon emission of silicon-vacancy centers located in different nanodiamonds. Our work enables ideal evanescent, optical coupling of distant nanodiamonds containing silicon-vacancy centers with applications in the realization of quantum networks, quantum repeaters or complex quantum systems.

\end{abstract}
%
%
\vspace{2pc}
\noindent{\it Keywords}: nanodiamonds, color centers in diamond, silicon-vacancy center, quantum optics, hybrid quantum systems
%
%
%
%

\section{Introduction}

Color centers in diamond promise to be the key element for upcoming quantum technologies such as quantum networks \cite{wehner2018quantum} or quantum sensors \cite{maze2008nanoscale, balasubramanian2008nanoscale}. The most studied color center so far is the negatively-charged nitrogen-vacancy (NV$^{-}$) center, due to its outstanding spin properties and magnetic field sensitivity \cite{jelezko2006single, doherty2013nitrogen}. For many applications, color centers close to the diamond surface are mandatory to allow efficient coupling to objects outside the diamond host. A special case are nanodiamonds (NDs), with dimensions small compared to the optical wavelength which, since recently, are also fabricated with high quality in top-down approaches \cite{zheng2019topdown}. In such cases, color centers inside the ND can be efficiently coupled to the outside optically via the evanescent field. Optical coupling gives access to, for example, post-processing of classical photonics towards quantum photonics applications \cite{fehler2019efficient, boehm2019onchip, schroeder2016quantum, benedikter2017cavity}. 
For efficient operation, it is important to optimize all degrees of freedom such as position, dipole alignment as well as the intrinsic optical properties, like a high flux of coherent photons, spectral stability and a narrow inhomogeneous distribution. The NDs offer the unique capability to optimize all degrees of freedom by means of atomic force microscope (AFM)-based nanomanipulation. In order to obtain an optimal evanescent coupling to an object outside the ND three basic operations are required. First, a high-precision translational manipulation \cite{schell2011ascanning}. Second, a high-precision rotational manipulation \cite{rogers2018single}. In case of NV$^{-}$ centers, the rotation angle was determined with respect to an external field by monitoring the change in ODMR signal. Third, the maintenance of ideal intrinsic properties. For optical coupling crucial intrinsic properties of the defect center include narrow homogeneous and inhomogeneous linewidths, good polarization contrast, low-background single photon emission and the ability of indistinguishable photon emission. The intrinsic properties remain a major drawback for NV$^{-}$ centers in NDs due to the degradation of the optical and spin properties when the NV$^{-}$ center is located in close proximity to the diamond surface. \\
In contrast, the negatively-charged silicon-vacancy (SiV$^{-}$) center has drawn great attention due to its robust intrinsic properties. In bulk diamond SiV$^{-}$ centers show exceptional optical properties \cite{hepp2014electronic, rogers2014electronic, sipahigil2014indistinguishable}, with a high Debye-Waller factor of $\sim 0.7$ \cite{dietrich2014isotopically}, a narrow inhomogeneous linewidth in the range of few GHz and Fourier-Transform limited homogeneous linewidth at cryogenic temperatures \cite{rogers2014multiple}. The spectral stability arises from the $\text{D}_{\text{3d}}$ symmetry of the SiV$^{-}$ center and enables to achieve excellent optical properties even in close proximity to the diamond surface. Protecting the high-degree of symmetry demands high quality of the diamond host, in particular, a low strain environment. In chemical vapour deposition (CVD) NDs single photon emission from SiV$^{-}$ centers \cite{neu2011single} and photostable quantum emitters with sub-gigahertz linewidth \cite{tran2017nanodiamonds} were shown. The production of low-strain nanodiamonds via HPHT fabrication method \cite{rogers2018single, bolshedvorskii2018single} recently enabled to preserve the excellent optical properties of SiV$^{-}$ centers also in NDs. In NDs with high crystal quality and sizes smaller than $200 \, \text{nm}$ an inhomogeneous distribution of $1.05 \, \text{nm}$ at $5 \, \text{K}$ and individual lines narrower than $360 \, \text{MHz}$ were found \cite{jantzen2016nanodiamonds}. Utilizing plasma treatment techniques of the diamond surface SiV$^{-}$ centers with an inhomogeneous ensemble linewidth below the excited state splitting and excellent spectral stability under resonant excitation was achieved \cite{rogers2018single}. While the requirement to maintain ideal intrinsic properties also in NDs has been shown the demonstration of nanomanipulation of NDs containing single SiV$^{-}$ centers is still missing. The translational nanomanipulation is a straightforward extension of the work done with NV$^{-}$ centers \cite{schell2011ascanning, rogers2018single}, but the rotational operation needs further developments. The determination of the rotation angle via ODMR contrast with respect to an external field, as done with NV$^{-}$ centers \cite{rogers2018single}, is not possible with SiV$^{-}$ centers. Furthermore, an open question remains on the fragility of the intrinsic properties on the nanomanipulation process. For example, conserving the intrinsic properties of the color centers during nanomanipulation on the NDs is crucial to enable the deterministic coupling of NDs with incorporated color centers with preselected properties, for example, in quantum photonics applications. \\
In this work we develop and test the complete post-processing procedure. We first isolate single SiV$^{-}$ centers in single ND, characterize the optical properties and choose the once with preferred optical properties comparable to SiV$^{-}$ centers in low-strain bulk diamond. We then perform the nanomanipulation operations, in particular translational and rotational operation needed for the post-processing step. Finally, we characterize the optical properties of the same SiV$^{-}$ centers again after nanomanipulation was performed and explicitly show that the optical properties of the color centers are conserved. The nanomanipulation enables optimization on all degrees of freedom of NDs hosting SiV$^{-}$ centers with an AFM cantilever. In addition we show that ND decomposition leads to a change in strain indicated by a change of the fine structure splitting. Finally, we proof the ability for indistinguishable, single photon emission from SiV$^{-}$ centers in different NDs.

\section{Methods}

Measurements presented in this manuscript were performed using a custom-built cryogenic confocal microscope. It consists of a compact design flow cryostat to cool down the sample to liquid helium temperature, a $\mathrm{NA} = 0.95$ microscope objective and a pinhole to achieve sub micrometer depth resolution. A dual-axis scanning galvo system is used to scan the sample surface, which is mounted on the cold finger of the cryostat. Off-resonant excitation of the color centers is done with a green DPSS laser at $\SI{532}{\nano\meter}$, while for resonant excitation we use a tunable diode laser at $\sim \SI{737}{\nano\meter}$. The fluorescence of the SiV$^{-}$ centers is collected with the microscope objective and analyzed either using a single photon counting module (SPCM) or a grating spectrometer with a $\SI{1800}{\mathrm{groves/mm}}$ grating. \\
NDs with SiV$^{-}$ color centers were obtained by high pressure – high temperature (HPHT) treatment of the catalyst metals-free hydrocarbon growth system based on homogeneous mixtures of naphthalene - $\mathrm{C}_{10}\mathrm{H}_{8}$ (Chemapol) and tetrakis(trimethylsilyl)sylane – $\mathrm{C}_{12}\mathrm{H}_{36}\mathrm{Si}_{5}$ (Stream Chemicals Co.), which was used as the doping component. Cold pressed tablets of the initial mixture ($\SI{5}{\milli\meter}$ diameter and $\SI{4}{\milli\meter}$ height) were placed into a graphite container, which simultaneously served as a heater of the high-pressure toroid-type apparatus. The experimental procedure consisted of loading the high-pressure apparatus to $\SI{8.0}{\giga\pascal}$ at room temperature, heating the sample to the temperature of diamond formation ($\sim \SI{1400}{\degree}\mathrm{C}$), and short ($1-3 \, \si{\second}$) isothermal exposure at this temperature. The recovered diamond materials have been characterized under ambient conditions by using X-ray diffraction, Raman spectroscopy and scanning and transmission electron microscopes (SEM and TEM). SEM images of the diamond nanoparticles can be found in the Supplemental Material. According to this characterization, diamond material, mainly, represent agglomerations relatively uniform nanoparticles that are $8-30 \, \si{\nano\meter}$ in size. In addition, samples may contain a small amount of larger fractions of diamond, which are formed in a narrow zone of direct contact of the original mixture with the inner surface of the graphite heater, corresponding to the area of maximum temperatures in the high-pressure apparatus. \\
The NDs were spin coated on a type-IIa-diamond substrate to ensure good thermal conductivity. After spin coating of the ND solution some of the NDs tend to cluster. A repeated deagglomeration could be achieved using an AFM cantilever as it is described in section 4. For better orientation the substrate is provided with a variety of markers that were produced by focused ion beam milling.

\section{Isolating single SiV$^{-}$ centers}

\begin{figure*}[hbtp]
	\includegraphics[width=0.96\textwidth]{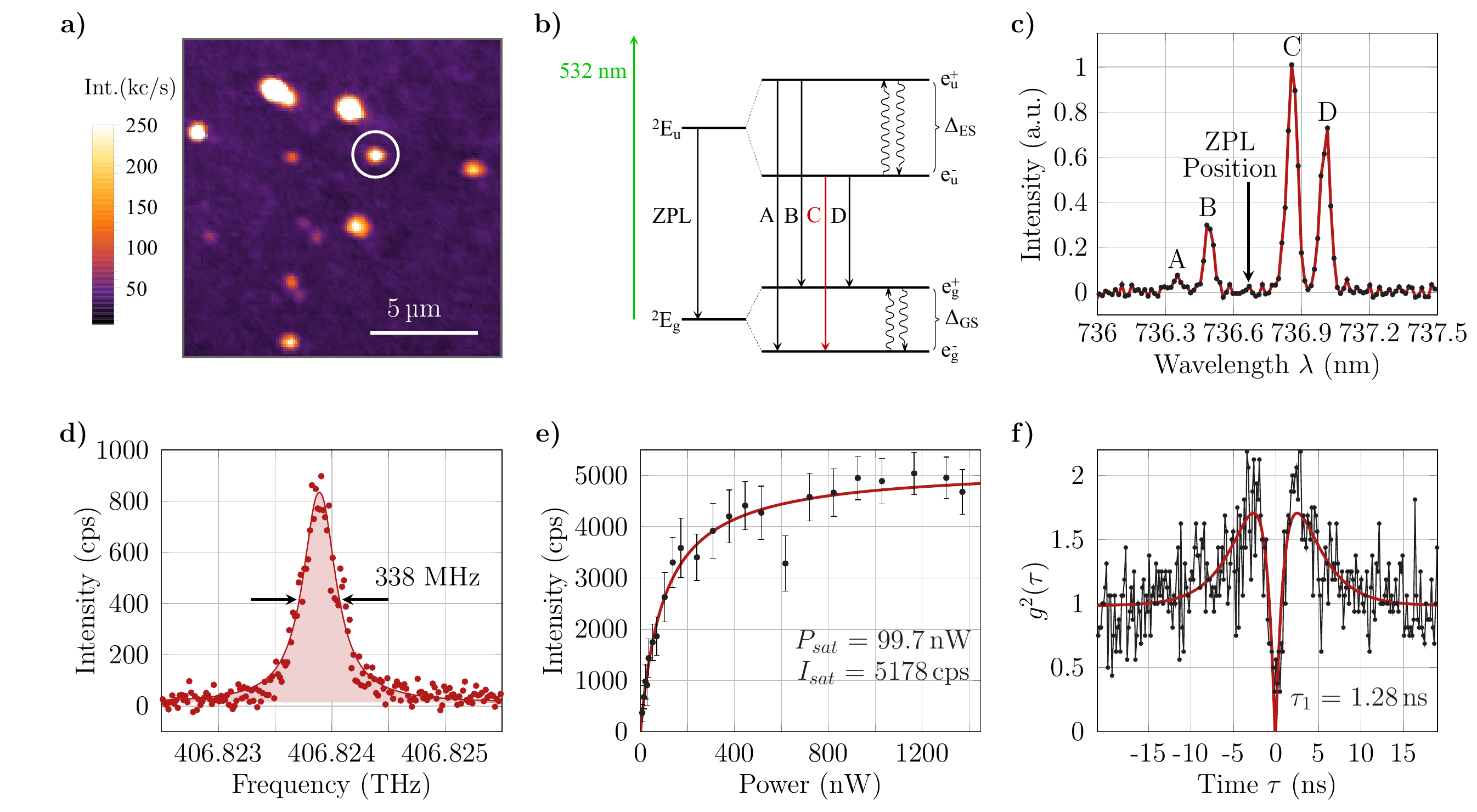}
		\caption{a) Confocal image of a $15 \times 15 \, \si{\micro\meter}^{2}$ region of the sample representing fluorescent spots with single SiV$^{-}$ centers (white circle). b) Levelscheme of the SiV$^{-}$ center.  c) PL spectrum of a single SiV$^{-}$ center in a ND. The four peaks (A-D) correspond to the four optical transitions of the SiV$^{-}$ ZPL (c.f. (b)). d) PLE scan of the transition C of an individual SiV$^{-}$ center at $\approx \SI{14}{\nano\watt}$ excitation power. The linewidth of $\SI{338 \pm 11}{\mega\hertz}$ indicates residual spectral diffusion (details see Supplemental Material). e) Saturation measurement of the SiV$^{-}$ center in (d) under resonant excitation resulting in a saturation power of $P_{\mathrm{sat}} = \SI{100 \pm 12}{\nano\watt}$ and a saturation intensity of $I_{\mathrm{sat}} = \SI{5178 \pm 150}{\mathrm{cps}}$ within the detection band of our $780/60$ bandpass filters. For the fit we used the model of equation (\ref{eqn:saturation}). f) g$^{2}$ measurement of the SiV$^{-}$ center in (d) at $\approx \SI{480}{\nano\watt}$ excitation power. The dip at $\mathrm{g}^{2}(0)$ clearly indicating single photon emission.}
	\label{fig:figure1}
\end{figure*}

The NDs occupy a low concentration of SiV$^{-}$ center potentially enabling to isolate single SiV$^{-}$ centers per one ND or cluster of NDs. In figure \ref{fig:figure1}(a) a confocal image of a $15 \times 15 \, \si{\micro\meter}^{2}$ area of the sample with several fluorescing spots is shown. Few of them show a four-line structure which corresponds to the four optical transitions (A-D) of the SiV$^{-}$ center zero-phonon line (ZPL) visible with off-resonant ($\SI{532}{\nano\meter}$) excitation (c.f. figure \ref{fig:figure1}(b)). The splitting of the electronic ground state ($\Delta_{\mathrm{GS}}$) and excited state ($\Delta_{\mathrm{ES}}$) is hereby caused by spin-orbit interaction. An exemplary low temperature photoluminescence (PL) spectrum of a SiV$^{-}$ center in a ND with its ZPL centered at $\approx \SI{736.7}{\nano\meter}$ is shown in figure \ref{fig:figure1}(c). A resolvable fine structure of single SiV$^{-}$ centers already indicates a high crystal quality and low strain of the NDs. The PL linewidth of the individual lines in figure \ref{fig:figure1}(c) is limited by the resolution of the grating spectrometer to about $\SI{20}{\giga\hertz}$. \\
Determination of the homogeneous linewidth of an individual SiV$^{-}$ center is done via photoluminescence excitation (PLE) spectroscopy (figure \ref{fig:figure1}(d)). We hereby scan the frequency of a resonant laser over transition C and record the fluorescence of the phonon sideband (PSB) using a $780/60$ bandpass filter. We measure a linewidth of $\SI{338 \pm 11}{\mega\hertz}$ under resonant excitation, about 3 times broader than the natural linewidth of SiV$^{-}$ that was reported in bulk diamond \cite{rogers2014multiple}. The broadening mainly arises from residual spectral diffusion. For the same ND we measure the powerdependence of the fluorescence under resonant excitation and fit a saturation curve following the model
\begin{equation}
I(P) = I_{\mathrm{sat}} P/P_{\mathrm{sat}} \frac{1}{1 + P/P_{\mathrm{sat}}}
\label{eqn:saturation}
\end{equation}
to the data (figure \ref{fig:figure1}(e)) yielding $P_{\mathrm{sat}} = \SI{100 \pm 12}{\nano\watt}$ and $I_{\mathrm{sat}} = \SI{5178 \pm 150}{\text{cps}}$. In a second-order autocorrelation measurement (figure \ref{fig:figure1}(f)) we find a $\mathrm{g}^{2}(0)$ value well below $0.5$ which implies that we indeed address a single SiV$^{-}$ center. \\
In the following we assume that we are investigating NDs containing a single SiV$^{-}$ center if three criteria are fulfilled. First, a well isolated fluorescent spot in confocal imaging. Second, a clear four-line structure in PL spectroscopy without any additional lines originating from a second SiV$^{-}$ center. Third, a single, Lorentzian-shaped PLE line scan under resonant excitation.
\newpage

\section{Translational nanomanipulation}

\begin{figure*}[hbtp]
	\includegraphics[width=0.96\textwidth]{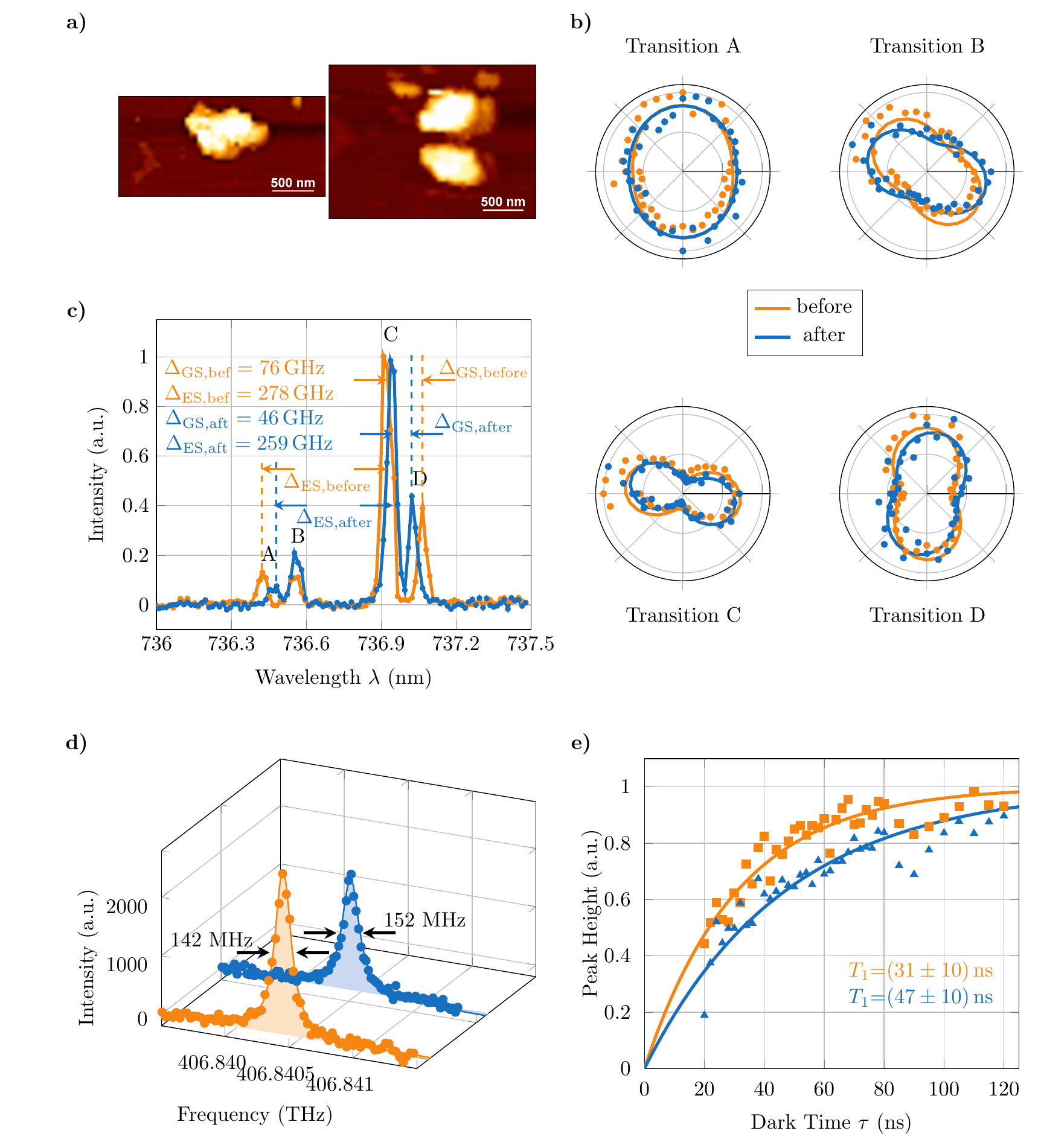}
		\caption{a) AFM image before (left) and after (right) the ND is displaced. The ND cluster is shifted and split into two parts on the right image. b) Polarization measurement (dots) of the four optical transitions (A-D) of the SiV$^{-}$ center and corresponding fits (solid lines) before (orange) and after (blue) the ND is displaced. The polarization contrast as well as the dipole angle of the individual transitions stay unchanged after the displacement of the ND. c) PL spectrum before (orange) and after (blue) the nanomanipulation. The four peaks (A-D) correspond to the four optical transitions of the SiV$^{-}$ ZPL. We observe a change of the ground state splitting from $\SI{76 \pm 4}{\giga\hertz}$ to $\SI{46 \pm 2}{\giga\hertz}$ and of the excited state splitting from $\SI{278 \pm 4}{\giga\hertz}$ to $\SI{259 \pm 2}{\giga\hertz}$, which indicates a decrease of the transverse strain of the ND after the nanomanipulation. d) PLE scan of transition C before (orange) and after (blue) the nanomanipulation at an excitation power of $\approx \SI{18}{\nano\watt}$. e) Orbital $T_{1}$ time measurement before (orange squares) and after (blue triangles) the nanomanipulation. We extract the $T_{1}$ time from an exponential fit using equation (\ref{eqn:t1time}).}
	\label{fig:translation}
\end{figure*}

In the following experiments we investigate the change of the optical properties of a ND containing a single SiV$^{-}$ center when it is moved with the cantilever of an AFM. An AFM image of the ND before and after the displacement is shown in figure \ref{fig:translation}(a).
Apparently from the size of the structure on the AFM image we conclude that we investigate a cluster of NDs. During the nanomanipulation the cluster is split in two parts and shifted on the substrate. The image quality of the second (shifted) image has significantly decreased due to a degeneration of the AFM cantilever when using the contact mode for the displacement of the ND. \\
We measure the emission polarization of the four optical transitions of the SiV $^{-}$ center (c.f. figure \ref{fig:figure1}(b)) at cryogenic temperatures before and after the nanomanipulation (see figure \ref{fig:translation}(b)). The lines A and D are perpendicular polarized compared to the lines B and C, which is known as the typical behavior of the SiV$^{-}$ center \cite{rogers2014electronic}. The polarization contrast as well as the dipole angle of the individual transitions stay unchanged after the displacement of the ND. We interpret this result as a translational movement without any rotation of the ND. The overall low polarization contrast can be explained by two main reasons. First, the polarization contrast of the individual lines highly depends on the orientation of the ND. For the A and D line, which can be described by a linear combination of X' and Y' dipoles (c.f. reference \cite{rogers2014electronic}), the polarization contrast changes if the ND is rotated around an angle $\Delta \theta$ with respect to the optical axis of the microscope objective (details see Supplemental Material). Second, the maximum polarization contrast is limited by the radiation characteristic of the SiV$^{-}$ center situated in the high refractive index ND into vacuum (n = 1). For a parallel (to the surface) oriented dipole the polarization contrast would be 1 while a perpendicular oriented dipole shows no polarization contrast. \\
We further investigate the ND properties before and after nanomanipulation via PL and PLE spectroscopy. The PL spectrum (figure \ref{fig:translation}(c)) shows the four line structure of the SiV$^{-}$ ZPL at $\approx \SI{737}{\nano\meter}$. After the displacement of the ND we observe a change of the ground state splitting $\Delta_{\mathrm{GS}}$ between the transitions A and B (C and D) from $\SI{76 \pm 4}{\giga\hertz}$ to $\SI{46 \pm 2}{\giga\hertz}$ and of the excited state splitting $\Delta_{\mathrm{ES}}$, which corresponds to the frequency difference between the transitions A and C (B and D), from $\SI{278 \pm 4}{\giga\hertz}$ to $\SI{259 \pm 2}{\giga\hertz}$. This can be explained by a decrease of the transverse strain when the ND is declustered. The splittings after the nanomanipulation are within the errorbars of the zero-strain splittings of $\Delta_{\mathrm{GS,ZS}} = \SI{46.3}{\giga\hertz}$ and $\Delta_{\mathrm{ES,ZS}} = \SI{252.0}{\giga\hertz}$ reported in reference \cite{rogers2018single}. The absolute line position of the ZPL stays unchanged, which we conclude with no additionally introduced longitudinal strain. \\
Resonant excitation of transition C before the nanomanipulation yields a linewidth of $\SI{142 \pm 3}{\mega\hertz}$ at an excitation power of $\approx \SI{18}{\nano\watt}$, which is close to the Fourier-Transform limit of a SiV$^{-}$ center in bulk diamond \cite{rogers2014multiple}. Repeating the measurement after the nanomanipulation yields a linewidth of $\SI{152 \pm 12}{\mega\hertz}$. In figure \ref{fig:translation}(d) both, the PLE scan before and after the ND was manipulated is shown. \\
Additionally, we measure the orbital relaxation time $T_{1}$ of the ground state via resonant laser pulses on transition C. During each laser pulse the fluorescence decays to a steady state level as the population is pumped to the other ground state by the resonant laser. The respective peak height depends on the dark time $\tau$ between the $\SI{200}{\nano\second}$ laser pulses. We compare the peak height of each pulse with the peak height of the first pulse and fit an exponential of the form
\begin{equation}
	h(\tau) = 1 - \exp(-\tau/T_{1})
\label{eqn:t1time}
\end{equation}
to the data, resulting in a $T_{1}$ time of $\SI{31 \pm 10}{\nano\second}$ ($\SI{47 \pm 10}{\nano\second}$) before (after) the nanomanipulation. The peak height data together with the exponential fit is depicted in figure \ref{fig:translation}(e). A slight increase of the $T_{1}$ time can be related to a better thermal conductivity after the displacement, which leads to a lower temperature of the ND and a longer orbital $T_{1}$ time of the SiV$^{-}$ center \cite{jahnke2015electron}. \\

\newpage

\section{Rotational nanomanipulation}

\begin{figure*}[hbtp]
	\includegraphics[width=0.96\textwidth]{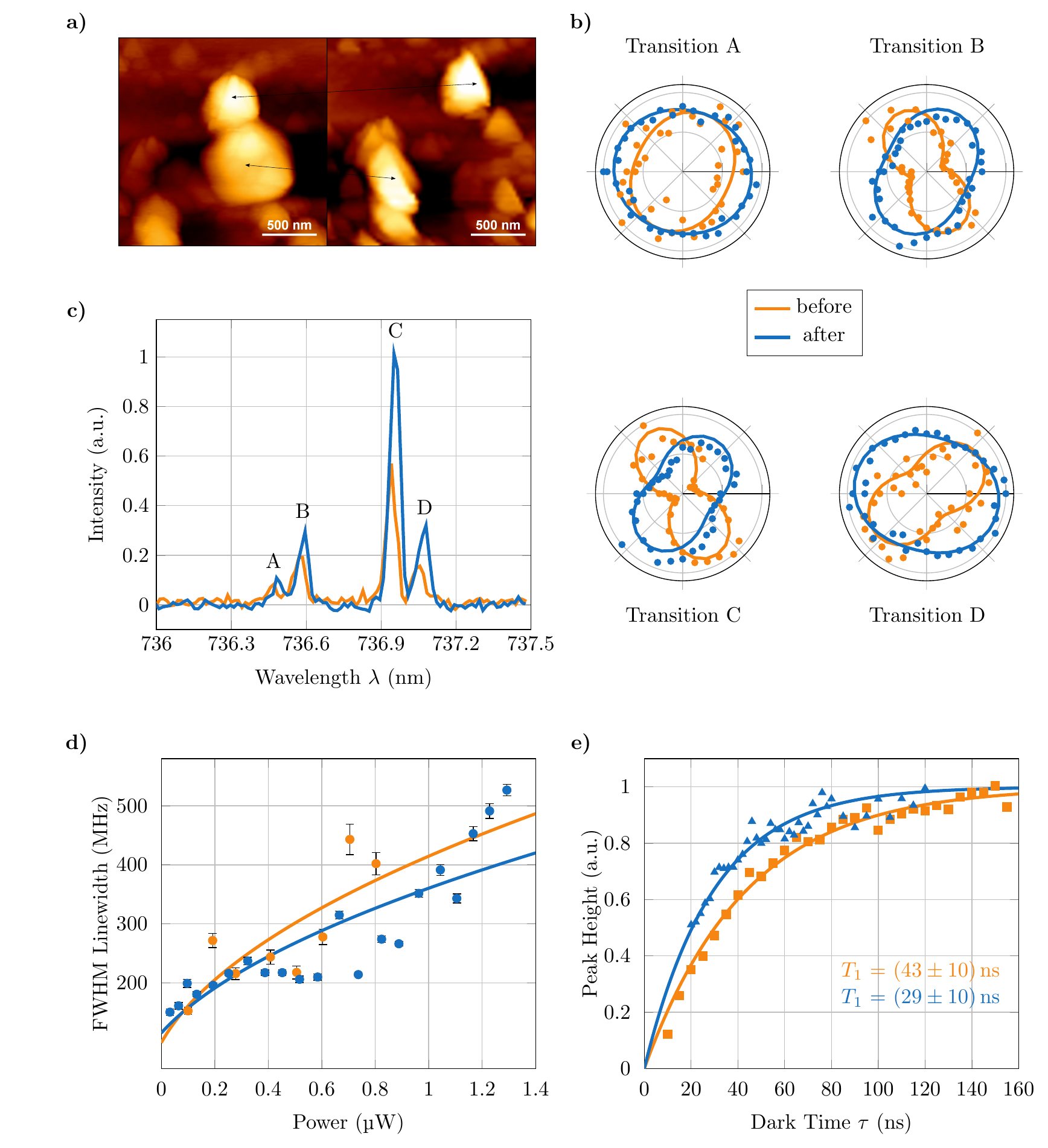}
		\caption{a) AFM image before (left) and after (right) the ND is displaced. The lower ND is moved $\approx \SI{1}{\micro\meter}$ to the left between the two images. Artifacts in the right image appear due to a degeneration of the AFM cantilever during contact mode. b) Polarization measurement (dots) of the four optical transitions (A-D) of the SiV$^{-}$ center and corresponding fits (solid lines) before (orange) and after (blue) the ND is displaced. The polarization angle is shifted by $\approx \SI{90}{\degree}$ after the nanomanipulation. c) PL spectrum before (orange) and after (blue) the nanomanipulation. Both, ground state and excited state splitting as well as the absolute line position stay unchanged, which implies that no additional strain is introduced to the ND during nanomanipulation. d) Power dependent PLE linewidth of transition C before (orange) and after (blue) the nanomanipulation. We measure a linewidth as narrow as $\SI{154 \pm 6}{\mega\hertz}$ ($\SI{151 \pm 6}{\mega\hertz}$) before (after) the nanomanipulation. The power dependency is described using equation (\ref{eqn:powerbroadening}). e) Orbital $T_{1}$ time measurement before (orange) and after (blue) the nanomanipulation. We extract the $T_{1}$ time from an exponential fit using equation (\ref{eqn:t1time}).}
	\label{fig:rotation}
\end{figure*}

Next, we demonstrate the rotation of a ND again using the AFM cantilever. An AFM image of the ND before and after the displacement is shown in figure \ref{fig:rotation}(a). The ND is laterally displaced by $\approx \SI{1}{\micro\meter}$ between the two images. \\
The rotation of the ND is verified by polarization measurements at cryogenic temperatures before and after the nanomanipulation (see figure \ref{fig:rotation}(b)).
From the rotation of the polarization axes of the transitions B and C during the nanomanipulation we extract the rotation of the ND in the x-y-plane, which is $\Delta \varphi = \SI{56 \pm 6}{\degree}$ for the investigated ND. Further we calculate the rotation of the ND around the angle $\Delta \theta$ via the change of the polarization contrast, which in our specific case is $\Delta \theta = \SI{15 \pm 6}{\degree}$ (details see Supplemental Material). \\
Beside the rotation of the polarization angle the optical properties of the SiV$^{-}$ center stay unchanged. A PL spectrum of the ND at cryogenic temperatures before and after the nanomanipulation is shown in figure \ref{fig:rotation}(c). We determine the ground state splitting of the SiV$^{-}$ center to $\SI{60 \pm 2}{\giga\hertz}$ ($\SI{61 \pm 3}{\giga\hertz}$) and the excited state splitting to $\SI{262 \pm 2}{\giga\hertz}$ ($\SI{263 \pm 3}{\giga\hertz}$) before (after) the nanomanipulation. This implies again a low strained ND, which holds true during the rotation. The overall brightness of the center has increased after the nanomanipulation due to better alignment of the SiV$^{-}$ dipole with the optical axis of the laboratory frame. \\
A comparison of the power-dependence of the PLE linewidth of transition C before and after the nanomanipulation is depicted in figure \ref{fig:rotation}(d). We again measure a narrowest linewidth close to the Fourier-Transform limit being $\SI{154 \pm 6}{\mega\hertz}$ before and $\SI{151 \pm 6}{\mega\hertz}$ after the nanomanipulation. The power broadening of the line is described by a model of the form
\begin{equation}
	\Delta\nu_{\text{FWHM}}(P) = \Gamma \left(1 + \frac{2 \Omega^{2}}{\Gamma^{2}} \right)^{1/2} = \left( \Gamma^{2} + \frac{2 \lambda^{3}}{\pi^{3} \hbar c d^{2}} \cdot \Gamma \cdot P \right)^{1/2},
\label{eqn:powerbroadening}
\end{equation}
where $\Omega$ is the Rabi Frequency, $\Gamma$ is the natural linewidth and $d$ is the beam diameter (see reference \cite{levine2012asimplified}). The behavior changes slightly during the rotation of the ND due to the change of the dipole alignment. \\
The measurement of the orbital relaxation time $T_{1}$ yields $\SI{43 \pm 10}{\nano\second}$ before and $\SI{29 \pm 10}{\nano\second}$ after the nanomanipulation, agreeing within the error margins. Both the measurement data and the exponential fit according to equation (\ref{eqn:t1time}) is shown in figure \ref{fig:rotation}(e). \\

\newpage

\section{Discussion and Conclusion}

\begin{figure*}[hbtp]
	\includegraphics[width=0.96\textwidth]{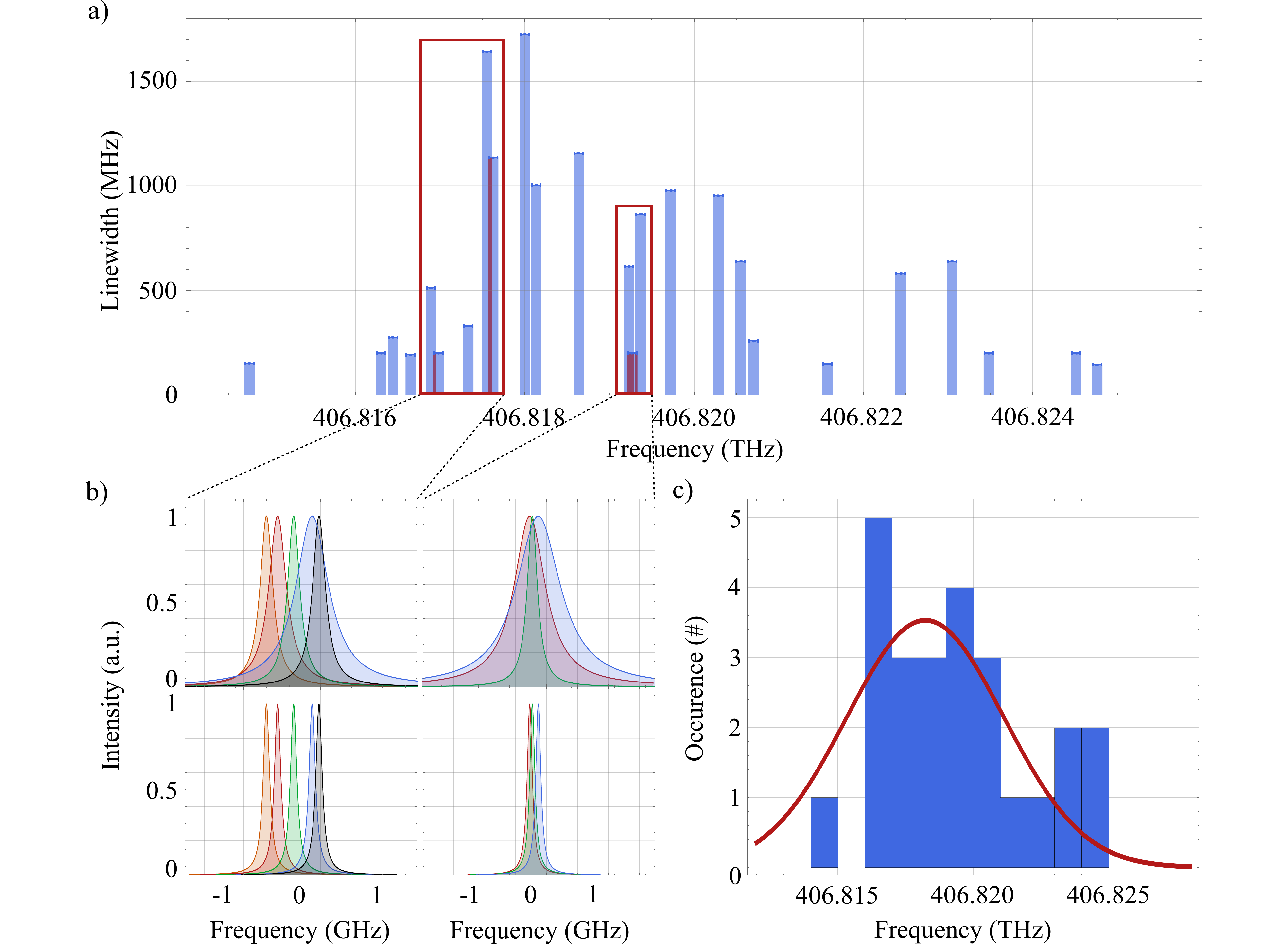}
		\caption{a) Linewidth versus center frequency of transition C of the 25 investigated SiV$^{-}$ centers. The width of the blue bars represents the natural linewidth of the SiV$^{-}$ center in bulk diamond ($\sim \SI{100}{\mega\hertz}$). Red areas indicate significant spectral overlap between individual centers. b) Lorentzian fits to the data, representing the measured center frequency of transition C and PLE linewidth of selected SiV$^{-}$ centers with spectral overlap (top). The actual data has been removed for clarity. We still see spectral overlap if we assume a linewidth of $\sim \SI{100}{\mega\hertz}$ (bottom). c) Spectral distribution of SiV$^{-}$ centers in the 25 investigated NDs. A Gaussian fit yields a FWHM of $\SI{6.8 \pm 0.9}{\giga\hertz}$.}
	\label{fig:figure4}
\end{figure*}

Summarizing, we have tested the complete post-processing procedure where one and the same SiV$^{-}$ center in a ND was pre-characterized, then nanomanipulation tools were applied on the ND host and finally the same SiV$^{-}$ center was characterized again. We have explicitly proven that the intrinsic properties of single SiV$^{-}$ centers with bulk-like optical properties persist against translational and rotational nanomanipulation of the ND host. Our work gives access to optical coupling of individual, pre-selected SiV$^{-}$ centers in NDs optimizing all degrees of freedom of the coupling term. In total we measured the ZPL position and linewidth (transition C) of 25 SiV$^{-}$ centers (figure \ref{fig:figure4}(a)).
We show significant spectral overlap of several SiV$^{-}$ centers in our NDs. 7 SiV$^{-}$ centers are promising candidates for a Hong-Ou-Mandel interference experiment as they could be assigned to a partner SiV$^{-}$ center with a detuning of the ZPL frequency smaller than the Fourier-Transform limited linewidth (red bars in figure \ref{fig:figure4}(a)). A zoom-in at 5 SiV$^{-}$ centers with an emission frequency of $\sim \SI{406.817}{\tera\hertz}$ in figure \ref{fig:figure4}(b) already demonstrates a small cluster with significant spectral overlap. Another cluster of 3 SiV$^{-}$ centers emits at $\sim \SI{406.819}{\tera\hertz}$ (figure \ref{fig:figure4}(b), right). In summary we find an inhomogeneous line distribution of $\SI{6.8 \pm 0.9}{\giga\hertz}$ of the 25 investigated SiV$^{-}$ centers (c.f. figure  \ref{fig:figure4}(c)), again indicating very low strain in our NDs. \\
With our results bottom-up approaches of coupled arrays of NDs with incorporated SiV$^{-}$ centers become feasible featuring the realization of quantum simulators. The deterministic coupling to photonic channels enables the construction of large-scale, distributed quantum networks. Also, magnetic field sensing at cryogenic temperatures could profit from well-controlled alignment of all degrees of freedom.

\newpage

\section*{Author Contributions}

SH, LH and AK performed the spectroscopic measurements. AFM measurements and nanomanipulation were performed by KGF. LFK, VAD and VNA synthesized the nanodiamonds with SiV$^{-}$ centers. The manuscript was written by SH, LH and AK and all authors discussed the results and contributed to the manuscript.

\section*{Acknowledgement}

The project was funded by the Deutsche Forschungsgemeinschaft (DFG, German Research Foundation) - Project number: 398628099. AK acknowledges support of the Carl-Zeiss Foundation, BMBF/VDI in project Q.Link.X and the Wissenschaftler-R\"uckkehrprogramm GSO/CZS. SH and AK acknowledge support of IQst. VAD and LFK acknowledge financial support of the Russian Foundation for Basic Research (Grant 18-03-00936). FJ acknowledges support from EU ASTERIQS (820394), the DFG (JE 290/18-1 and SFB 1279), BMBF/VDI in project Q.Link.X and ERC (BioQ 319130).
We thank Gregor Neusser, Christine Kranz and the FIB Center UUlm for etching the markers into the type-IIa-diamond substrate. Experiments performed for this work were operated using the Qudi software suite \cite{binder2017qudi}.

\nocite{*}
\section*{References}
\bibliographystyle{unsrt}
\bibliography{Preparing_single_SiV_center_in_nanodiamonds}

\newpage

\title[SM for Preparing single SiV$^{-}$ center in nanodiamonds for external, optical coupling]{Supplemental Material for "Preparing single SiV$^{-}$ center in nanodiamonds for external, optical coupling with access to all degrees of freedom"}

\author{Stefan H\"au{\ss}ler \textsuperscript{\bfseries 1,2},
  Lukas Hartung \textsuperscript{\bfseries 1},
  Konstantin G. Fehler \textsuperscript{\bfseries 1,2},
  Lukas Antoniuk \textsuperscript{\bfseries 1},
  Liudmila F. Kulikova \textsuperscript{\bfseries 3},
  Valery A. Davydov \textsuperscript{\bfseries 3},
  Viatcheslav N. Agafonov \textsuperscript{\bfseries 4},
  Fedor Jelezko \textsuperscript{\bfseries 1,2} and
  Alexander Kubanek \textsuperscript{\bfseries 1,2}}

\address{\textsuperscript{1} Institute for Quantum Optics, Ulm University, Albert-Einstein-Allee 11, 89081 Ulm, Germany \\
  \textsuperscript{2} Center for Integrated Quantum Science and Technology (IQst),
Ulm University, Albert-Einstein-Allee 11, 89081 Ulm, Germany \\
  \textsuperscript{3} L.F. Vereshchagin Institute for High Pressure Physics,
Russian Academy of Sciences, Troitsk, Kaluzhskoe shosse 14, Moscow 142190, Russia \\
  \textsuperscript{4} GREMAN, UMR CNRS CEA 6157, Université F. Rabelais, Parc de Grandmont, 37200 Tours, France}
  
\ead{alexander.kubanek@uni-ulm.de}

\vspace{10pt}

\begin{indented}
\item[]October 2019
\end{indented}

%
%

%
\vspace{2pc}
%
%
%
%

\section*{Determination of the Rotation Angle}

\begin{figure*}[hbtp]
		\begin{center}
			\includegraphics[width=0.72\textwidth]{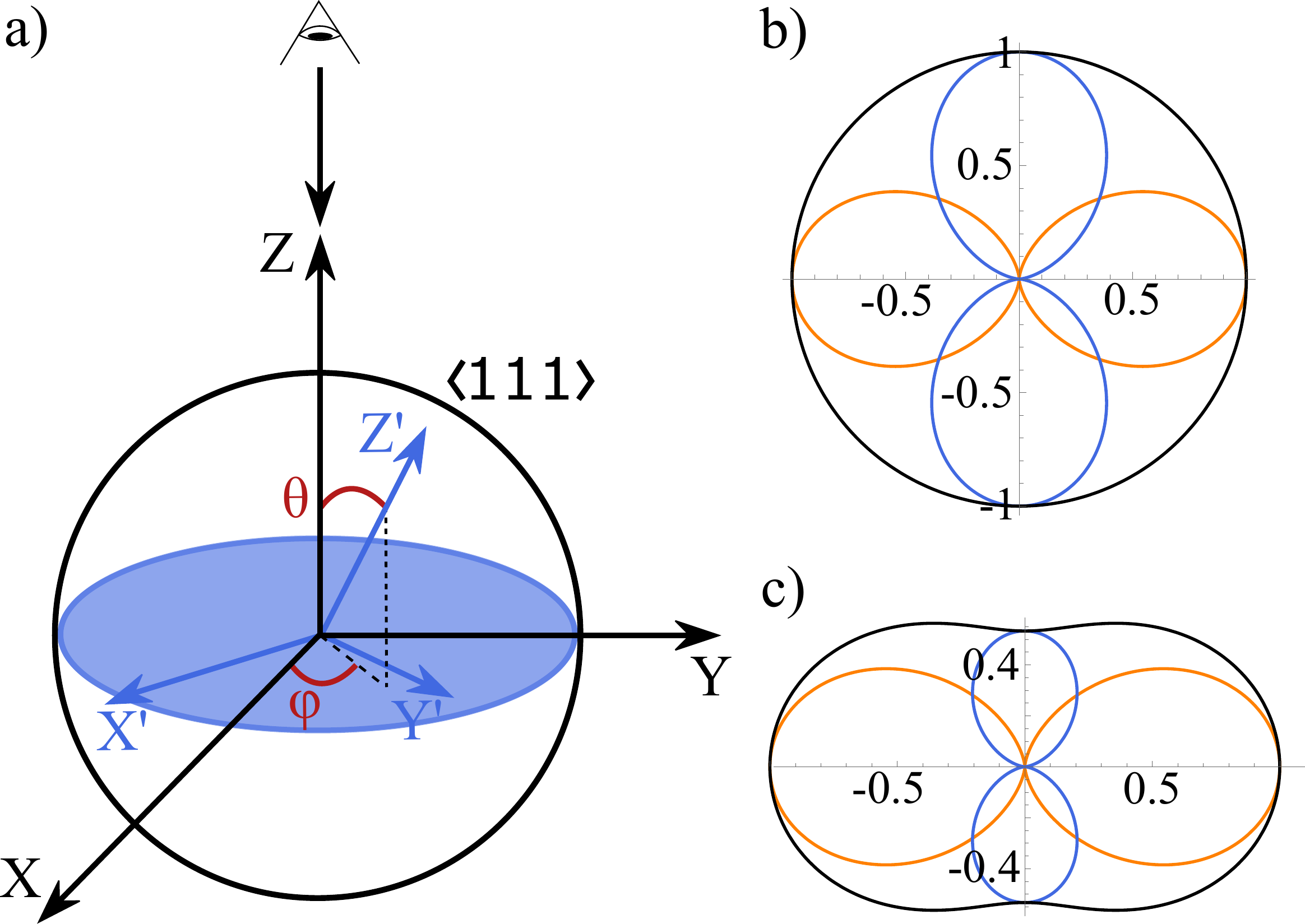}
				\caption{a) Lab frame $X$, $Y$, $Z$ and optical dipoles $X'$,$Y'$ and $Z'$ of a randomly oriented ND with the relative angles $\theta$ and $\phi$. b), c) Projection of the $X'$ (orange) and $Y'$ (blue) dipole emission on the x-y-plane for $\theta = \SI{0}{\degree}$ (b) and $\theta = \pi/4$ (c). The sum of the two dipoles (black curve) illustrates a change of the polarization contrast.}
			\label{fig:rotationangle}
		\end{center}
\end{figure*}

We determine the rotation angles $\Delta \varphi$ and $\Delta \theta$ (figure \ref{fig:rotationangle}) of the dipoles of the SiV$^{-}$ center during the nanomanipulation via the rotation of the polarization axes in the polarization plots and the change of the polarization contrast
\begin{equation}
C = \frac{I_{\mathrm{max}} - I_{\mathrm{min}}}{I_{\mathrm{max}} + I_{\mathrm{min}}},
\label{eqn:polcontrast}
\end{equation}
where $I_{\mathrm{max}}$ ($I_{\mathrm{min}}$) are the maximum (minimum) intensity in the polarization measurement (c.f. reference \cite{rogers2014electronic}). \\
For a given lab frame $X$, $Y$, $Z$ a randomly oriented SiV$^{-}$ center can be fully described by three orthogonal dipoles $X'$, $Y'$, $Z'$ with the angles $\varphi$ and $\theta$. Here the $Z$ axis denotes the optical axis of our microscope objective (viewing direction) while the $Z'$ axis is chosen along the $\left< 111 \right>$ direction of the ND lattice. This means that for the transitions B and C the only active dipole is the $Z'$ dipole \cite{rogers2014electronic}. A rotation of the polarization axes of these two transitions can therefore be attributed to a rotation of the ND around the angle $\Delta \varphi$. We measure a change of the polarization angle of transition B from $\SI{115}{\degree}$ to $\SI{63}{\degree}$ and of transition C from $\SI{117}{\degree}$ to $\SI{56}{\degree}$ during the nanomanipulation. We therefore conclude that the ND was rotated around the angle $\Delta \varphi = \SI{56 \pm 5}{\degree}$. \\
The transitions A and D can be described by a linear combination of the dipoles $X'$ and $Y'$ (c.f. reference \cite{rogers2014electronic}). A rotation of the ND around the angle $\Delta \theta$ therefore leads to a change of the polarization contrast in the polarization plots of the transitions A and D (see figure \ref{fig:rotationangle}(b),(c)). The polarization contrast of the A and D transitions can be calculated using equation (\ref{eqn:polcontrast}):
\begin{equation}
C = \frac{1 - \cos^{2} \theta}{1 + \cos^{2} \theta}.
\label{eqn:polcontrastad}
\end{equation}
For the angle $\theta$ we obtain
\begin{equation}
\theta(C) = \arccos \left( \sqrt{\frac{1 - C}{1 + C}} \right).
\label{eqn:theta}
\end{equation}
From our measurements we calculate a change of the polarization contrast from $\si{13}{\, \%}$ to $\si{6}{\, \%}$ for transition A and from $\si{40}{\, \%}$ to $\si{12}{\, \%}$ for transition D during the nanomanipulation, resulting in a rotation of the ND around the angle $\Delta \theta$ = \SI{15 \pm 6}{\degree}. \\
\newpage

\section*{ND fabrication}

\begin{figure*}[hbtp]
		\begin{center}
			\includegraphics[width=0.72\textwidth]{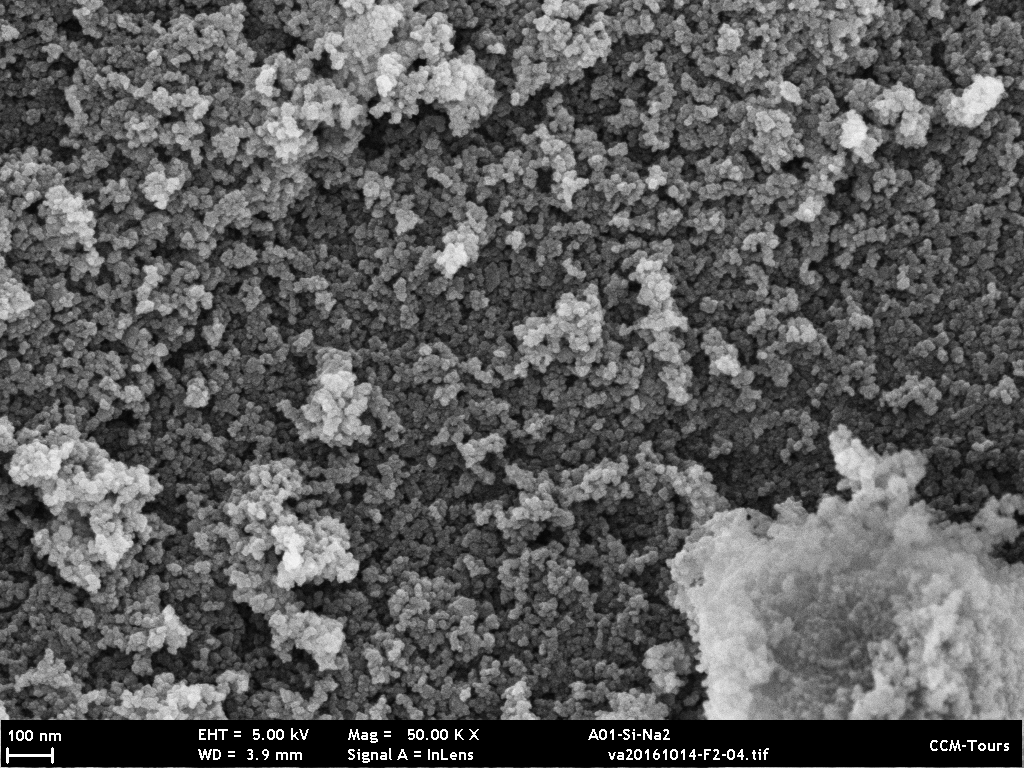}
			\includegraphics[width=0.72\textwidth]{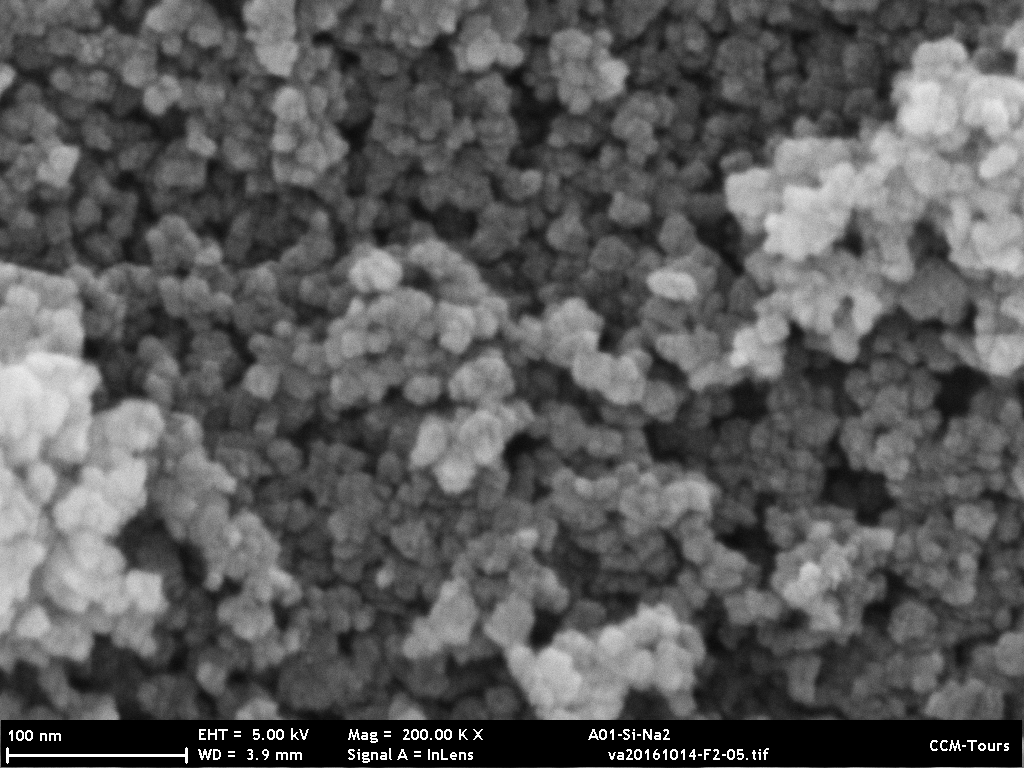}
				\caption{SEM images of diamond nanoparticles with SiV$^{-}$ color centers obtained by high pressure – high temperature (HPHT) treatment of the catalyst metals-free hydrocarbon growth system based on homogeneous mixtures of $\mathrm{C}_{10}\mathrm{H}_{8}$ and $\mathrm{C}_{12}\mathrm{H}_{36}\mathrm{Si}_{5}$.}
			\label{fig:semimages}
		\end{center}
\end{figure*}

The details of the ND synthesis are described in the main text. We here show SEM images of the diamond nanoparticles that are $\approx 8-30 \, \si{\nano\meter}$ in size (figure \ref{fig:semimages}).
\newpage

\section*{Evaluation of spectral diffusion in PLE scans}

\begin{figure*}[hbtp]
		\begin{center}
			\includegraphics[width=0.76\textwidth]{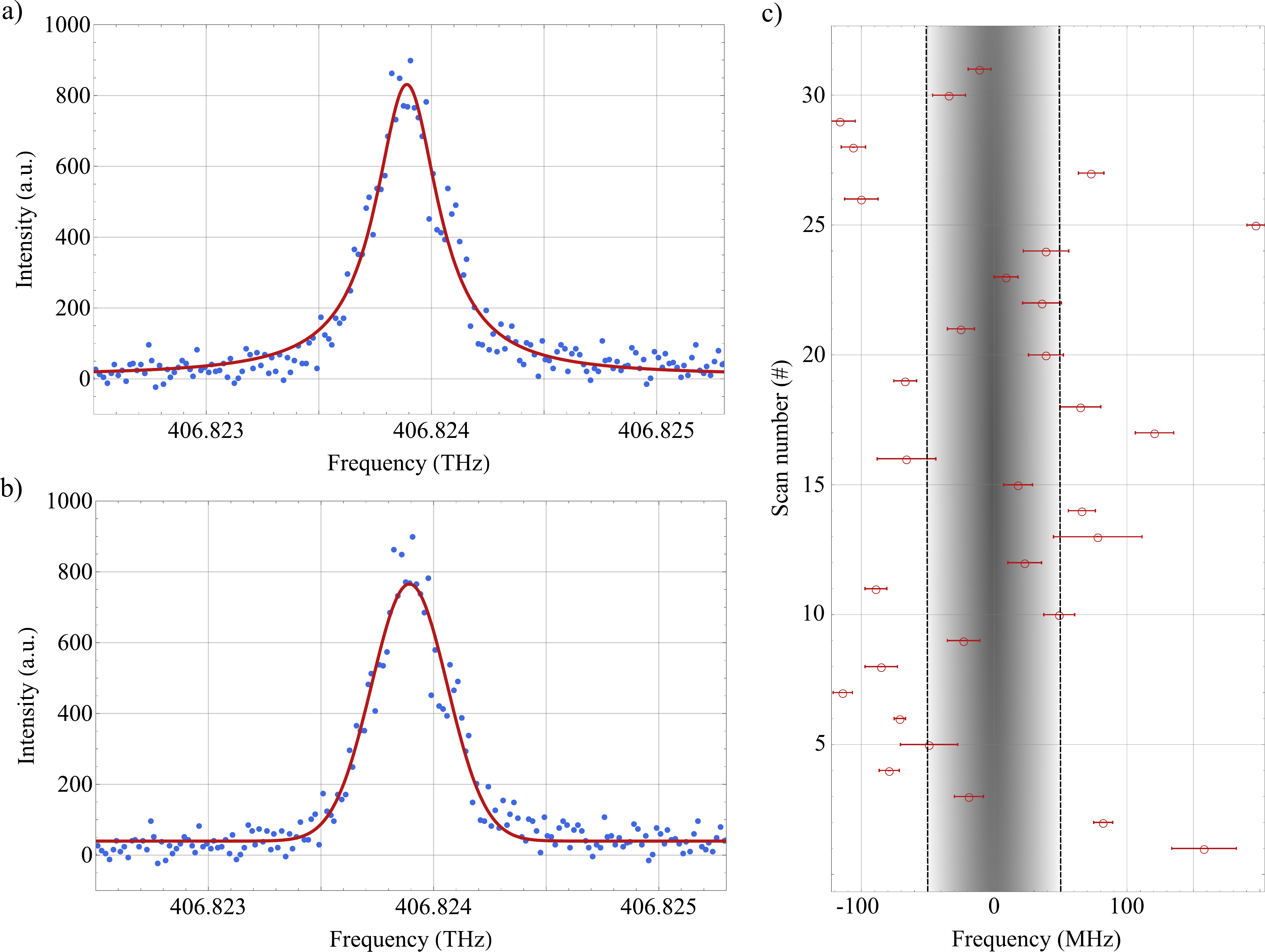}
				\caption{a) Histogram of PLE scans of transition C of the SiV$^{-}$ center of figure 1(d) of the main paper at $\approx \SI{14}{\nano \watt}$ excitation power. The linewidth of $\SI{338 \pm 11}{\mega \hertz}$ is here extracted from a Lorentzian fit. b) Same histogram data. The Gaussian fit yields a linewidth of $\SI{396 \pm 9}{\mega \hertz}$. The fit underestimates the peak maximum resulting in a broader linewidth compared to the Lorentzian fit. c) Center frequency for single PLE scans showing spectral jumps, which can be interpreted as spectral diffusion. The grey bar represents the Fourier-Transform limited linewidth of the SiV$^{-}$ center in bulk diamond.}
			\label{fig:diffusion}
		\end{center}
\end{figure*}

We study spectral diffusion of transition C of the SiV$^{-}$ center of figure 1(d) of the main paper. Therefore we plot the histogram of several PLE scans and fit the data with a Lorentzian function (figure \ref{fig:diffusion}(a)) and a Gaussian function (figure \ref{fig:diffusion}(b)). The Gaussian fit underestimates the peak maximum resulting in a broader linewidth compared to the Lorentzian fit. In the main paper we therefore extract the PLE linewidth of the SiV$^{-}$ center from the Lorentzian fit. \\ 
Spectral diffusion of the line becomes visible when investigating the peak position of single PLE scans (figure \ref{fig:diffusion}(c)). We observe spectral jumps of the resonance frequency on the order of the previously determined histogram linewidth of $\SI{338}{\mega \hertz}$.

\end{document}